\documentclass[twocolumn,prl,preprintnumbers,amsmath,amssymb]{revtex4}
\usepackage{amsmath,dcolumn,graphicx,epsf,ulem}
\setkeys{Gin}{width=8.6cm,keepaspectratio}
\usepackage{dcolumn}
\usepackage{bm}
\usepackage{color}
\usepackage{nicefrac}

\begin{document}

\title{Evolution of magnetic oxygen states in Sr-doped LaCoO$_3$}

\author{S. Medling$^1$,  Y. Lee$^2$, H. Zheng$^3$, J.~F. Mitchell$^3$, J.~W. Freeland$^4$, B.~N. Harmon$^2$, F. Bridges$^1$} 

\affiliation{$^1$Department of Physics, University of California,
 Santa Cruz, California 95064, USA}

\affiliation{$^2$Ames Laboratory and Iowa State University,
Ames IA 50011-3020}

\affiliation{$^3$Materials Science Division, Argonne National
Laboratory, 9700 Cass Ave, Argonne, IL 60439, USA}

\affiliation{$^4$Advanced Photon Source, Argonne National
Laboratory, 9700 Cass Ave, Argonne, IL 60439, USA}

\date{\today}

\begin{abstract}

Magnetism in La$_{1-x}$Sr$_x$CoO$_3$ as a function of doping is investigated
with X-ray absorption spectroscopy (XAS) and X-ray magnetic circular dicrhoism
(XMCD) at the O K-edge, and corresponding first principles electronic structure
calculations. For small $x$, the spectra are consistent with the formation of
ferromagnetic clusters occurring within a non-magnetic insulating matrix.
Sr-induced, magnetic O-hole states form just above E$_F$ and grow with
increasing Sr doping.  Density functional calculations for $x$=0 yield a
non-magnetic ground state with the observed rhombohedral distortion and indicates
that doping introduces holes at the Fermi level in magnetic states with significant
O 2p and Co t$_{2g}$ character for the undistorted pseudo-cubic structure.
Supercell calculations show stronger magnetism on oxygen atoms having more Sr
neighbors.

\end{abstract}


\maketitle

Doping in complex oxides is a fundamental route for control of the many
degrees of freedom. However, depending on the nature of the underlying
ground-state, the impact on the local electronic structure depends strongly on
the balance between on-site energies and the bonding to
oxygen\cite{Imada:1998p38738,Goodenough:2004p34845}. For example, doping in the
Mott regime directly changes the valence state of the transition metal ion
(\textit{e.g.}\ manganites) while doping in the extreme of the charge transfer regime
(\textit{e.g.}\ cuprates) results in formation of holes on the oxygen lattice. The case
of cobaltites lies in the cross-over region between these two extremes and a
better understanding of this regime may finally answer the
decades-old question of why magnetism in La$_{1-x}$Sr$_{x}$CoO$_3$ (LSCO) is so
unusual.

LaCoO$_3$, is a non-magnetic, small gap semiconductor at low T,
but at increased temperature and particularly with Sr doping
on the La sites, the material exhibits a wealth of unusual magnetic properties
associated with Co sites acquiring a substantial magnetic moment ($\sim$1.3$\mu_B$).
For $x$ $>$ 0.18 the crystal becomes
ferromagnetic (FM) and metallic,\cite{Itoh94,AnilKumar98,Sikolenko04,Aarbogh06,He07}---see
Fig. 5 of Ref.  \onlinecite{He07} for a phase diagram.   With increasing
$x$, the rhombohedral structural distortion gradually decreases and the
crystal approaches the cubic perovskite structure\cite{Kriener:2009p7662}.

Most studies of LSCO have  
used localized models which assume that the dominant interactions are the
crystal field splitting and the exchange interactions E$_{ex}$; these
interactions split the Co 3d energy levels into narrow t$_{2g}$ and e$_g$
multiplets and also split the spin up and spin down states.  Within such models
there is extensive debate about the spin states, with low spin (S=0,
t$_{2g}^6$e$_g^0$)\cite{Yamaguchi96} at low T and a possible mixture of low spin and
either an intermediate spin (S=1, t$_{2g}^5$e$_g^1$)\cite{Knizek:2005p38727}
or a high spin state (S=2, t$_{2g}^4$e$_g^2$) at higher T or with Sr doping.
However, we argue that this localized picture is not appropriate---the e$_g$
states interact with O 2p states and are spread over a large energy range (10
eV), from low energy bonding states to high energy anti-bonding states.

In Co K-edge X-ray absorption near edge spectroscopy (XANES) 
studies\cite{Sundaram09,Jiang09} the Co K-edge shifts $\leq$ 0.15 eV as $x$
increases to 0.3, in stark contrast to other Co systems and many manganites,
for which the edge shift is roughly 3 eV per valence
unit\cite{Sikora06,Han08,Bridges01}.  Further, both diffraction
studies\cite{Caciuffo99,Mineshige99} and extended X-ray absorption fine structure (EXAFS) results\cite{Jiang09} show that
the Co-O bond length is nearly independent of $x$ for LSCO; consequently, the
bond-valence sum model also indicates no change in Co valence---\textit{i.e.} the 3d
electron configuration remains close to d$^6$. This raises the question: where
do the Sr-induced holes go?  We argued that a large fraction of the holes are
on the O atoms\cite{Jiang09}, which can be directly probed at the oxygen K
edge. Previous XAS studies at the O K-edge show an evolution with hole
doping\cite{Saitoh97,Toulemonde01}, but existing X-ray magnetic circular dichroism
(XMCD) studies are limited.\cite{Okamoto00,Merz10}
Thus, to date there has been no systematic study of the low hole doping regime to
follow the connection  between local structure and changes in electronic and magnetic states.

Here, we present a detailed doping-dependent study of the oxygen hole states in
LSCO
single crystals, using polarization-dependent XAS to track the first O pre-edge
peak (Co-O ligand hole states). It is clear that a magnetic state with a large
O component, forms with doping and the increase in O absorption is tied to a
large fraction of the doped holes residing on the oxygen site. The strong XMCD
shows directly that there is a non-zero orbital moment on oxygen (and
correspondingly, a net O magnetic moment).
The analysis shows the moment on oxygen is parallel to the Co moment and
increases in magnitude with doping. 
To understand these results and determine the magnetic structure of O,
complementary theoretical studies have been carried out; they show that the 3d
electron density of states (DOS) are spread over a large energy range, with
significant e$_g$ character for bonding states 6 eV below E$_F$. They also show
that the (empty) O DOS just above the Fermi level increases with $x$, and that
the O atoms develop a moment, with both spin and orbital components.  Most
importantly, the O atoms are no longer equivalent; their properties depend on
the number of nearest Sr neighbors. 

\begin{figure}

\includegraphics[width=3.0in]{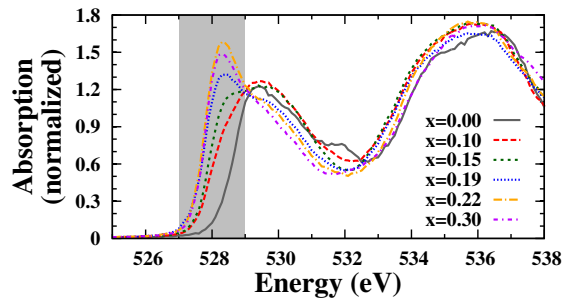}

\caption{XAS at the O K-edge. The XAS
data were normalized at 550 eV and self-absorption corrections were applied.
The data agree well, except at the beginning of the edge (from
527-529 eV) where, with increasing $x$, a large
increase in the XAS is observed, corresponding to the addition of empty states.}
\vspace{-0.1in}
\label{O-xas}
\end{figure}

XMCD and XAS data for the O K-edges were measured at the Advanced Photon Source
beamline 4-ID-C. A superconducting magnet provided
magnetic fields up to $\pm$5T. The energy resolution was $\sim$ 0.15 eV.
Small single crystals were mounted on an
electrode using Ag paste and scraped with a diamond file to remove surface contamination.
The X-ray beam was oriented at 45$^{\circ}$ to the sample surface and the
fluorescence detector was at 90$^{\circ}$ to the beam.

\begin{figure}

\includegraphics[width=3.0in]{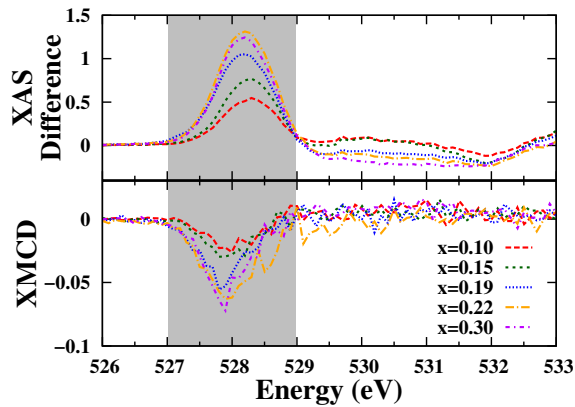}

\caption{Difference in the XAS (top) and XMCD (bottom) at the O K-edge.
A significant XMCD signal is only observed over this same small energy range (shaded region)
where change in the XAS occurs and XAS($x$)--XAS(0) is positive.
Above 529 eV, the differences become slightly negative.}
\vspace{-0.1in}

\label{O-diff}
\end{figure}

The XAS fluorescence data at the O K-edge
are shown in Fig.\ \ref{O-xas}. A small constant background was subtracted from the XAS
and then the data were normalized to 1 at energy 550 eV.  Finally, the data were corrected for self
absorption following the method in the computer program FLUO\cite{Fluo}.
The main results for the XAS
data as $x$ increases are a shift of the leading part of the O K-edge to lower
energy and an increase in absorption. As has been noted
previously\cite{Sarma:1996p38789} and in our theory results below, the weak
core-hole interaction at the O K-edge means that the XAS closely
follows the shape of the O-projected density of states (DOS).  This allows us
to make a direct connection with the evolution of electronic structure.
 
The magnetic state on O is probed by XMCD ($\mu^+$--$\mu^-$, where $\mu^+$ and
$\mu^-$ are the absorption coefficients for $\pm$ circularly polarized X-rays
with the sample in a B-field high enough that the magnetization is close to
saturation) as shown in the lower part of Fig.\ \ref{O-diff}.  A significant
signal is only observed over a short energy range ($\sim$ 2 eV) at the
beginning of the O K-edge, indicating a non-zero orbital moment on the O
site\cite{Igarashi:1994p38784}.  The amplitude of the XMCD signal increases
with $x$ and is surprisingly large above $x$ = 0.19 compared to measurements of
O XMCD in other systems; the formation of the FM metallic state appears to
arise from the decreasing rhombohedral distortion with $x$.  However, the fact
that only a small part of the O XAS shows a magnetic signal indicates that not
all the O hole-states are magnetic.

To visualize how the electronic structure evolves with $x$, the
Sr-induced change in XAS defined as XAS($x$) -- XAS($x$ = 0) is shown in the top of
Fig.\ \ref{O-diff}.  Sr induces a strong increase
in the absorption at lower energy due to the formation of additional
unoccupied oxygen ligand hole states. For low $x$, this new XAS
peak shifts to lower energies and increases in amplitude as $x$ increases,
with no other changes in the XAS up to $\sim$ 533 eV.  For $x$ $>$ 0.15,
there is no further shift, but the amplitude continues to increase with $x$ and
the XAS difference becomes negative above 529 eV (\textit{i.e.} the O-DOS decreases from
529-532 eV above the metal-insulator transition (MIT)).
The shaded regions in Figs.\ \ref{O-xas} and \ref{O-diff} illustrate
that the energy range for which the XMCD is observed corresponds to the
same energy range over which the additional peak in the XAS develops. 

\begin{figure}
\includegraphics[width=3.0in]{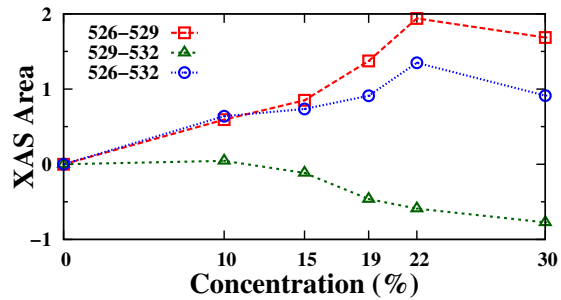}

\caption{Integrals of the difference in O XAS (Fig.\ \ref{O-diff}) from
526-529 eV (magnetic holes) and from 529-532.5 eV (non-magnetic holes) as a
function of $x$; note that the change is not linear. For $x$
$\leq$ 0.15, the new magnetic holes are associated with FM clusters; over the
range 0.15-0.22 where the material becomes fully FM, higher energy holes with
no orbital momentum are converted to lower energy magnetic states.}
\vspace{-0.1in}

\label{integ}
\end{figure}

This increase in empty O-DOS is consistent with Sr doping adding holes to the
hybridized O bands.  Note, however, that the density of the oxygen states with
quenched angular momentum (i.e. above 529 eV) decreases with $x$ for $x$ above
$\sim$ 0.18.  The induced O magnetic holes (DOS integral from 526-529 eV) and
non-magnetic holes (from 529-532 eV) are shown in in Fig.\ \ref{integ}.  There
is a clear change in these integrals near $x$ = 0.18 where the MIT occurs and
the system changes from a spin glass to a FM state.

To understand this evolution, we utilize doping-dependent electronic structure calculations.
Recent studies suggest that a first principles density functional
approach from an itinerant electron viewpoint can provide insights about the
behavior of LSCO.\cite{Takahashi:1998p38792,Ravindran99,Ravindran02}
This is a homogeneous approach, which is typically applicable for
itinerant or free-electron systems.  We use the electronic states obtained
from first principles band structure calculations to calculate theoretical XAS
and XMCD spectra. For the electronic structure, we
use a full-potential linearized augmented plane-wave (FPLAPW)
method\cite{Blaha01} with a local density functional.\cite{Perdew92, Perdew96}
We used R$_{MT}$K$_{max}$ = 8.0 and R$_{MT}$ = 2.2, 2.2, 1.9 and 1.6 a.u.  for
La, Sr, Co and O respectively.

\begin{figure}
\vspace{0in}
\includegraphics[width=1.9in,angle=0]{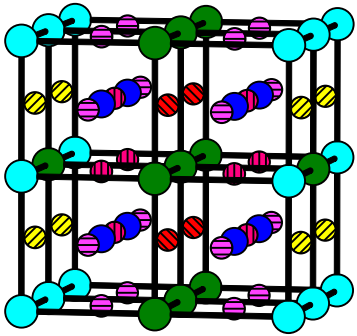}

\caption{2$\times$2$\times$2 supercell for a Sr doping of 3/8.
Sr (large, light blue); La (large, dark green); Co (medium, blue).  Four types of (small)
O atoms with varying Sr neighbors: 0 (red, left slant); 1 (magenta, vertical);
2 (pink, horizontal); 3 (yellow, right slant). }
\vspace{-0.1in}

\label{cell}
\end{figure}

Because of the strong short range electronic interactions, the virtual crystal
approximation (VCA) is not appropriate.  Consequently, we use a
2$\times$2$\times$2 supercell formed of eight pseudo-cubic unit cells, with Sr
concentrations 0, 0.125, 0.25 and 0.375 as shown in Fig.\ \ref{cell}.  We
initially kept the cubic structure even for the lower concentrations, since
preliminary calculations indicated that the spectral features were more
sensitive to the local atomic arrangement compared to the overall cell geometry.  For
the supercells, 40 k-points were used for the iterations to self-consistency.

\begin{figure}
\includegraphics[width=3.4in,angle=0]{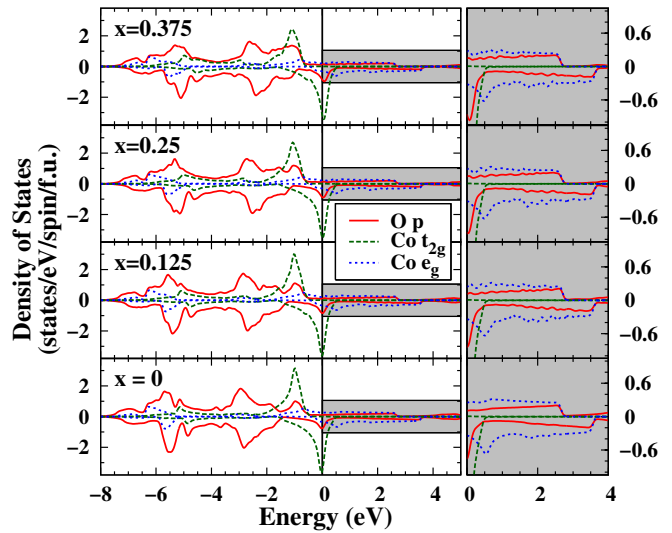} 

\caption{Orbitally decomposed DOS for Co 3d (t$_{2g}$ and e$_g$) and O 2p
states from LDA calculations. For each panel top is spin up, bottom, spin down.
The shaded region for E$\geq$E$_F$ is expanded on right.  }

\label{DOS} 
\end{figure}

The strong interactions between O 2p and Co 3d (particularly e$_g$) orbitals
form well-separated bonding (--6 eV)  and antibonding (--1 to 4 eV) states---see
the orbital decomposed DOS in Fig.\ \ref{DOS}. A local model with e$_g$ mainly
above the t$_{2g}$ states is inconsistent with these DOS. The generalized
gradient approximation (GGA) and local density approximation (LDA)
density functionals yield very similar DOS and only the LDA results are
plotted.  Our calculations for undoped LaCoO$_3$ are essentially in agreement
with those of Ravindran {\it et al.}\cite{Ravindran99,Ravindran02}.  However,
although the GGA\cite{Perdew96} is more sensitive to the electronic
correlations and yields a nonmagnetic ground state with experimental
rhombohedral distortion (Co-O-Co bonding angle is 163.4$^{\circ}$) LDA needs
stronger distortion (Co-O-Co bonding angle is 158.4$^{\circ}$) to yield the
nonmagnetic state.  The transition from the nonmagnetic to magnetic ground
state (with $\sim$1.3$\mu_B$ moments on Co) is not unlike the traditional
local transition state picture, in the sense that about one Co d electron is
promoted from spin down to spin up with very little cost in energy, and is very
sensitive to the rhombohedral distortion.
Using the experimental value for the rhombohedral distortion, the GGA calculation
shows a $\sim$ 3.2 meV/Co total energy difference between non-magnetic and
magnetic states.

\begin{figure}
\includegraphics[width=3.0in,angle=0]{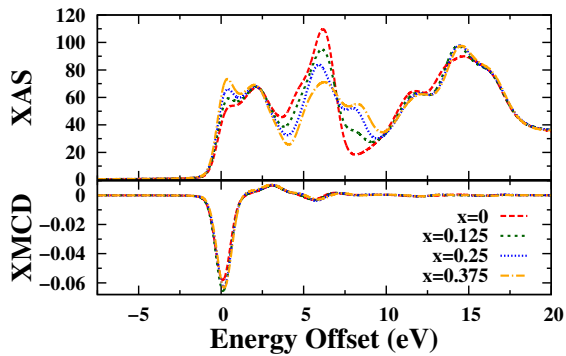}

\caption{Theoretical O K-edge (top), showing a significant increase at the
beginning of the edge, and calculated O K-edge XMCD signal (bottom).}

\vspace{-0.1in}

\label{XAS}
\end{figure}

We also extracted the theoretical O K-edge (mainly a measure of the oxygen
partial DOS (PDOS)) and the O XMCD signal, both as a function of $x$, as seen
in Fig.\ \ref{XAS}.  These theoretical results agree quite well with the
experimental results shown in Figs.\ \ref{O-xas} and \ref{O-diff}, though it is
important to keep in mind that the energies are relative to $E_F$, so the XAS
does not shift to lower energy as is observed experimentally, but the increases
are quite similar and, again, we see there is only a large XMCD signal over the
same energies for which the XAS has a large increase.  The lack of an amplitude
change with $x$ is likely because the rhombohedral distortion was not included
for small $x$.

\begin{figure}

\includegraphics[width=3.0in,angle=0]{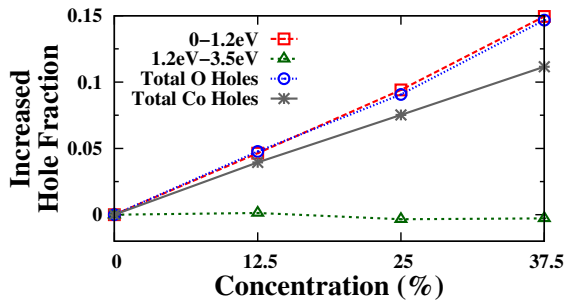}

\caption{Calculated hole fractions: magnetic O-hole fraction (red
square), non-magnetic O-hole fraction (green triangle), total O-holes (blue
circle), total Co-holes (gray star).}
\vspace{-0.1in}

\label{holes}
\end{figure}

Further important details are obtained by calculating the Sr-induced hole
fraction on Co and O atoms, using similar energy ranges as for Fig.\
\ref{integ}.  In Fig.\ \ref{holes}, we plot the induced hole density as a
function of $x$; the calculated O hole density is comparable to but slightly
larger than that on the Co atoms. Thus, surprisingly the induced holes are nearly equally
distributed over the Co and O atoms, in good agreement with the experimental
findings.

The calculations of the individual O magnetic moments are even more unusual and
show that the O atoms in the supercell are not equivalent.  Their properties depend
on the number of neighboring Sr atoms, which can vary within the supercell
(from 0 to up to 3 for $x$=0.375): the larger the number of Sr neighbors, the larger the
calculated magnetic moment (from 0.067$\mu_B$ for O with 0 Sr neighbors to
0.121$\mu_B$ for those with 3). Thus, these calculations
show the limitations of the VCA; the system is not homogeneous at the local
level. A similar situation occurs for Co, showing different Co moments on different sites,
when calculated using a trigonal distortion of the basic unit cell, since the cubic 2$\times$2$\times$2
supercell is too small to discriminate between different Co.
 Another interesting feature from the calculated O-PDOS (Fig.\
\ref{DOS}) is that it provides a simple explanation as to why there are
magnetic and non-magnetic holes; just above E$_F$ the spin up and spin down
O-PDOS are not equal so those energy states will have a net moment. At higher
energies the spin up and down O-PDOS are nearly equal and will have little
moment.

In summary, we find that it is better to view these cobaltites in terms of
bonding and anti-bonding states rather than split t$_{2g}$ and e$_g$ states.
Some of the e$_g$ symmetry states form bonding states and are filled;
consequently, there is never the situation where some e$_g$ are empty. A mixture
of filled e$_g$ and t$_{2g}$ states may explain the wide range of conflicting
experimental results about the Co spin state. Further, the broad range of the
e$_g$ DOS precludes a Jahn-Teller (J-T) interaction from being important. This likely explains the
lack of a significant J-T distortion for the Co-O bonds.\cite{Sundaram09,Jiang09}

The experimental changes in the O XAS and XMCD are not linear with $x$ and
appear to be correlated with the overall magnetism in the system, which changes
from glass-like with FM clusters below 18\% to metallic FM above 18\%.
Calculations qualitatively agree quite well with both the XAS and XMCD at the O
K-edge, and indeed find a small but significant spin magnetic moment on O
atoms.  Further, the theoretical results show that the properties of the O
atoms are not all equivalent, and have different moments; confirming that the
VCA is not adequate for describing systems such as LSCO, for which the electron
density and oxygen moments are sensitive to nearby Sr positions.  With part of
the doped holes being on the O atoms, the Co-O Coulomb attraction that usually
contributes to a shorter metal-O bond when holes are added, is reduced and
likely is a major reason as to why there is so little change in the Co-O bond
length with doping. 

Note: a recent explosive high field measurement up to 500 T on single crystal
LaCoO$_3$ indicated a magnetic state appearing above 140 T with a magnetic moment
of $\sim$ 1.4 $\mu_B$, which could correspond to the excited state with the same
moment obtained by our band structure calculations.\cite{Platonov12}

The XAS and XMCD measurements were carried out at the Advanced Photon Source at
Argonne National Laboratory and is supported by the U.S. Department of Energy,
Office of Science under grant No. DE-AC02-06CH11357.  Work at the Ames
Laboratory was supported by the US Department of Energy, Basic Energy Sciences
under grant No.  DE-AC02-07CH11358.



\end{document}